\documentclass[a4paper,twoside]{article}

\usepackage{epsfig}
\usepackage{subcaption}
\usepackage{calc}
\usepackage{url}
\usepackage{amsfonts}
\usepackage{bm}
\usepackage{amssymb}
\usepackage{amstext}
\usepackage{amsmath}
\usepackage{amsthm}
\usepackage{multicol}
\usepackage{pslatex}
\usepackage{apalike}
\usepackage{aas_macros}
\usepackage{SCITEPRESS}     

\usepackage[pdftex,colorlinks=true,breaklinks=true,citecolor=blue,draft=false]{hyperref}

\DeclareMathOperator*{\argmin}{argmin}

\begin{document}

\title{Sparse Image Reconstruction for the SPIDER Optical Interferometric Telescope}

\author{\authorname{Luke Pratley\sup{1, 2}\orcidAuthor{0000-0002-4716-9933} and Jason D. McEwen\sup{2}\orcidAuthor{0000-0002-5852-8890}}
\affiliation{\sup{1}Dunlap Institute for Astronomy and Astrophysics, \\ University of Toronto, Toronto, ON M5S 3H4, Canada}
\affiliation{\sup{2}Mullard Space Science Laboratory (MSSL), University College London (UCL), Holmbury St Mary, Surrey RH5 6NT, UK}
\email{luke.pratley@utoronto.ca, Jason.McEwen@ucl.ac.uk}
}

\keywords{Optical, Photonic Circuits, Interferometry, Imaging, Convex Optimization}

\abstract{The concept of a recently proposed small-scale interferometric optical imaging device, an instrument known as the Segmented Planar Imaging Detector for Electro-optical Reconnaissance (SPIDER), is of great interest for its possible applications in astronomy and space science. Due to low weight, low power consumption, and high resolution, the SPIDER telescope could replace the large space telescopes that exist today. Unlike traditional optical interferometry the SPIDER accurately retrieves both phase and amplitude information, making the measurement process analogous to a radio interferometer. State of the art sparse radio interferometric image reconstruction techniques have been gaining traction in radio astronomy and reconstruct accurate images of the radio sky. In this work we describe algorithms from radio interferometric imaging and sparse image reconstruction and demonstrate their application to the SPIDER concept telescope through simulated observation and reconstruction of the optical sky. Such algorithms are important for providing high fidelity images from SPIDER observations, helping to power the SPIDER concept for scientific and astronomical analysis.}

\onecolumn \maketitle \normalsize \setcounter{footnote}{0} \vfill

\section{\uppercase{Introduction}}
\label{sec:introduction}

\noindent The Hubble Space Telescope (HST) has changed the way astronomers have looked at the Universe. The number of astronomical studies that have used observations from the HST make it one of the most important observatories in history. More than 15,000 articles that have used HST data, in total collecting 738,000 citations.\footnote{See \url{https://www.nasa.gov/mission_pages/hubble/story/index.html}} However, telescopes such as the HST and its scientific successor, the James Webb Space Telescope (JWST), are extremely heavy and large, while being expensive in cost and power consumption. Nevertheless such next generation optical telescopes like JWST are critical to address astronomy and cosmology science goals such as answering questions about dark matter through weak lensing and understanding the history and formation of our universe.

Recently, the concept of an instrument known as the Segmented Planar Imaging Detector for Electro-optical Reconnaissance (SPIDER) has been developed \cite{ken13,dun15}.
The SPIDER is a small-scale interferometric optical imaging device that first uses a lenslet array to measure multiple interferometer baselines, then uses photonic integrated circuits (PICs) to miniaturize the measurement acquisition. The goal of the \mbox{SPIDER} is to reduce the weight, cost, and power consumption of optical telescopes. Furthermore, additional designs have been proposed that could increase the efficiency of imaging using fewer measurements \cite{liu17,liu18}. Recent visibility measurements using lenslet arrays and PICs have shown to match theoretical predictions \cite{su17}. Unlike traditional optical interferometry, the SPIDER telescope can accurately retrieve both phase and amplitude information \cite{su17}, making the measurement process analogous to a radio interferometer. Accurate interferometric image reconstruction methods from radio astronomy can thus be applied to image SPIDER observations.

Radio astronomy has a long history of using interferometry to push beyond the limits of resolution and size, at the computational cost of image reconstruction \cite{ryl60}. An interferometer is a device that measures the cross-correlation function of the signals. Interferometric imaging in the radio has proven to be a popular approach between 50 MHz and 100 GHz, with telescopes such as the Very Large Array (VLA) that have antenna arrays spread over 36 kilometers \cite{tho08}. The cross-correlation between voltages from each pair of antenna is computed to interfere the complex valued measurements known as visibilities. A visibility represents a Fourier coefficient for the sky brightness, with the Fourier coordinate determined by the antenna pair separation. Typically an antenna pair is known as a baseline, with the baseline length corresponding to the antenna separation \cite{tho08}.

Recently, sparse image reconstruction algorithms that exploit developments from the field of convex optimization have shown to improve the quality of reconstructed observations from radio interferometers considerably, on both simulations and real data \cite{pra18,cai17b,pra18a,pra19c,pra19e}. In this article we take recent developments from radio interferometric imaging and sparse image reconstruction, and put them into the context of the proposed SPIDER instrument. Such methodology would prove useful in future space based telescopes and space missions based on the SPIDER technology (e.g.\ aerial observations of planetary surfaces). Ultimately it is evident that recent algorithmic developments for radio interferometric imaging can be directly applied to the SPIDER optical interferometer.

In Section \ref{sec:spider} we introduce the background and current developments behind the SPIDER concept. Section \ref{sec:measurement_operator} discusses the calculation of the SPIDER measurement equation. Section \ref{sec:sparse_regularization} introduces the sparse image reconstruction formalism. 
Section \ref{sec:reconstructions} shows image reconstruction from a simulated SPIDER observation. 

\section{SPIDER}
\label{sec:spider}
Key to the concept design of the SPIDER is the use of lenslets to collect signals from incoming light.  These signals are combined using a PIC to produce an interferometric measurement (visibility), i.e.\ a Fourier coefficient of the observation. The Fourier coordinates, $(u, v)$, are determined by the separation size in wavelengths (baseline length) between the lenslets that were used to generate the measurement, with larger separations resulting in higher resolution measurements.
However, unlike radio interferometry where all possible pairs of antennas in an array can be combined in an observation, lenslets can only be paired once. If there are $N_{l}$ lenslets, the lenslet array will produce $N_l/2$ correlations.  This differs to the $N(N - 1)/2$ correlations expected from a radio array \cite{tho08,liu18}. To compensate for this lenslets can be combined with the PIC to split the signal into spectral bins (channels), allowing for increased sampling coverage due to variation of baseline length over wavelength. This strategy has been successful in radio astronomy for decades, and is known as multi-frequency synthesis \cite{tho08}.

\begin{figure}
    \center
    \includegraphics[width=0.4\textwidth]{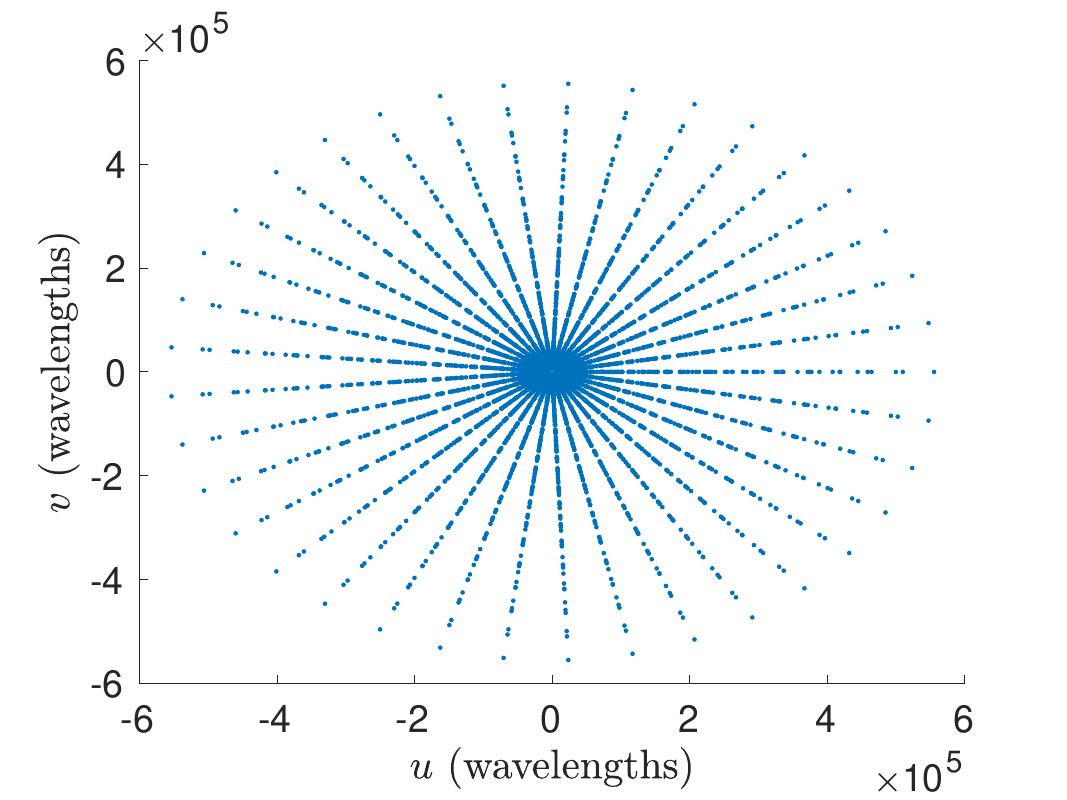}
    \caption{The sampling pattern of SPIDER in the $uv$-plane in units of wavelengths using 24 lenslets over 37 PIC cards for the combined coverage of 10 spectral bins. The sampling pattern was generated using the parameters in Table \ref{tab:spider_specs}. Since the Fourier coordinates are relative to wavelength, using the spectral bins (channels) will increase the 
    $uv$-coverage of the instrument substantially. The number of measurements in the single channel corresponds to 444, which makes 4440 measurements over the entire band.}
    \label{fig:uvcoverage}
\end{figure}

The concept design of the SPIDER proposed in \cite{ken13} is to put a linear array of lenslets onto a PIC card. The PIC cards are mounted as radial spokes on a disc, producing a radial sampling pattern in the $uv$-plane (however, other sampling patterns are considered in \cite{ken13}). The proposed operating wavelengths are between 500 nm and 900 nm. The operating wavelength divided by the size of a lenslet (8.75 mm) determines the field of view to be approximately between 0.5 and 1 arc minutes. The longest baseline along a spoke is 0.5 m, which is sensitive to resolutions between 0.65 and 1.2 arcseconds. Parameters of the SPIDER design adopted from \cite{ken13} are listed in Table \ref{tab:spider_specs}, which leads to the $(u, v)$ sampling coverage shown in Figure \ref{fig:uvcoverage}.

\begin{table}
\center
\caption{SPIDER configuration parameters adopted from \cite{ken13}.}
\label{tab:spider_specs}
\begin{tabular}{l|c}
\hline
Parameter & Value \\
\hline
Spectral Coverage    &   500-900 nm\\
Lenslet Diameter & 8.75 mm\\ 
Longest Baseline & 0.5 m \\
Number of Lenslets per PIC spoke & 24 \\
Number of PIC spoke & 37 \\
Number of Spectral Bins &10 \\
FoV at 500 (900) nm &  35$^{\prime\prime}$  (65$^{\prime\prime}$)\\
Maximum Resolution at 500 (900) nm & $~$ 0.7$^{\prime\prime}$ (1.2$^{\prime\prime}$)\\
Total Measurements & 4440\\
\hline
\end{tabular}
\end{table}

\section{SPIDER Measurement Model}
\label{sec:measurement_operator}
The SPIDER is an interferometric instrument operating at optical wavelengths. It follows from \cite{zer38} that the measurement equation will have the form of a Fourier Transform
\begin{equation}
  y(u, v) = \int_{-\infty}^\infty x(l, m) a(l, m) {\rm e}^{-2\pi i (l u + m v)}{\rm d}l {\rm d}m\, ,
\end{equation}
where $(l, m)$ is the coordinates of the image intensity $x$, $a$ is the sensitivity of the instrument over the field of view, and $(u, v)$ is the separation vector between two lenslets and the Fourier coordinate of the measurement \cite{tho08}. Due to the limited number of lenslets it is not possible to sample the entire $(u, v)$ domain, leading to an ill-posed inverse problem. However, the measurement equation is linear, and we can represent the measurement equation as a linear operator
\begin{equation}
\label{eqn:inv-pro}
  \bm{y} = \mathsf{\bm{\Phi}} \bm{x} + \bm{n}\, ,
\end{equation}
with measurements $\bm{y}_k = y(\bm{u}_k, \bm{v}_k)$ for a sampling pattern $(\bm{u}, \bm{v})$ and intensity values $\bm{x}_i = x(\bm{l}_i, \bm{m}_i)$ for pixels of an image $(\bm{l}, \bm{m})$ and $\bm{n}$ is independent and identically distributed (i.i.d.) Gaussian noise. 

The measurement operator requires performing a Fourier transform. However, the sampling pattern of SPIDER does not lie on a regular grid so it is not possible to use the speed of the standard Fast Fourier Transform (FFT). Instead, a non-uniform FFT can be used to evaluate the measurement equation, where an FFT is applied, and the visibilities are interpolated off of the FFT grid. This is known as the non-uniform FFT (NUFFT) but is commonly known as degridding in radio astronomy \cite{fes03,tho08}. 

In a discrete setting, let $\bm{x} \in \mathbb{R}^{N}$ and $\bm{y} \in \mathbb{C}^{M}$. Due to the limited field of view for $\bm{x}$, the values of $\bm{y}$ can be estimated using interpolation off the FFT grid. Sinc interpolation suppresses repeating artifacts from outside the imaged region appearing in the model $\bm{x}$. These artifacts are known as aliasing error, this becomes more difficult to suppress for bright sources outside the field of view. However, Sinc interpolation requires convolution of the FFT grid with a Sinc function for each measurement $\bm{y}_i$, which requires a large support. In general, it is possible to replace the Sinc function with an interpolation kernel with minimal support such as a Kaiser-Bessel or Gaussian kernel \cite{fes03}, with appropriate apodization correction in the image domain.

A NUFFT can be represented by the following linear operations 
\begin{equation}
  \mathsf{\bm{\Phi}} = \bm{\mathsf{W}}\bm{\mathsf{G}}\bm{\mathsf{F}}\bm{\mathsf{Z}}\bm{\mathsf{S}} \, ,
  \label{eq:matrix_equation}
\end{equation}
where $\bm{\mathsf{S}}$ represents scaling required due to the attenuation of the interpolation kernel but also scaling due to the sensitivity of the instrument over the field of view, $\bm{\mathsf{Z}}$ is a zero padding operation of the image to up sample the Fourier domain, $\bm{\mathsf{F}}$ is an FFT, $\bm{\mathsf{G}}$ represents the convolution constructed from the interpolation kernel used to interpolate measurements off of the grid, and $\bm{\mathsf{W}}$ are noise weights applied to the measurements, i.e $\bm{\mathsf{W}}_{kk} = 1/\sigma_{k}$ for a variance of $\sigma_{k}^2$. 
This measurement operator is described in detail in \cite{pra18} where it is applied to radio interferometric imaging. 
$\mathsf{\bm{\Phi}}$ has its adjoint operator $\mathsf{\bm{\Phi}}^\dagger$, which consists of applying adjoints of these operators in reverse. In this work we assume that $\bm{y}$ are the weighted measurements and that the dirty map is defined as $\mathsf{\bm{\Phi}}^\dagger \bm{y}$.

\subsection{Interpolation kernels}
\label{sec:kernels}
The sparse degridding matrix $\bm{\mathsf{G}}$ can be constructed in 1d from a kernel $d(u)$ by 
\begin{equation}
  \bm{\mathsf{G}}_{i, \{k_i + j\}_K} = d(u_i - (k_i + j)) \, ,
\end{equation}
where $i$ is the index of the measurement $y_i$, $k_i$ is the closest integer to visibility coordinate $u_i - J/2$ (in units of pixels) and is found by the flooring operation, and $j = 1 \dots J$ are the possible non-zero entries of the kernel. $\{ \cdot \}_K$ is the modulo-$K$ function, where $K = \alpha \sqrt{N}$ is the dimension of the Fourier grid in 1-D (for notational sake, the 2d Fourier grid is comprised of $K\times K$ samples). For a separable anti-aliasing kernel, it is straight forward to extend the 1d kernel to 2d. The diagonal operator $\bm{\mathsf{S}}$ is calculated in a similar way by
\begin{equation}
  \bm{\mathsf{S}}_{i,i} = s\left(\frac{i}{K} - \frac{1}{2}\right)\, ,
\end{equation}
where $s(x)$ is the reciprocal of the inverse Fourier transform of $d(u)$. In practice, $\bm{\mathsf{S}}$ can be computed analytically if the inverse Fourier transform of the convolution kernel is known. 

The Kaiser-Bessel kernel has shown to perform well as an interpolation kernel while using a minimal support size \cite{fes03,pra18}. More examples of interpolation kernels are provided in \cite{jac91} and \cite{tho08}.  For this work we use the Kaiser-Bessel kernel that was used in \cite{pra18}.

\section{Sparse Image Reconstruction}
\label{sec:sparse_regularization}
The Bayesian statistical inference framework can be used to address the inverse problem \eqref{eqn:inv-pro}, as shown in \cite{cai17a}. From Bayes' theorem,
the posterior can then be expressed as
\begin{equation}
  p\left(\bm{x} | \bm{y} \right) \propto \exp\bigl[ -\|\bm{y} - \mathsf{\bm{\Phi}} \bm{x}\|_{\ell_2}^2/2\sigma^2\bigr] \exp\bigl[ -\gamma \| \bm{\mathsf{\Psi}}^\dagger \bm{x}\|_{\ell_1}\bigr]\, ,
\end{equation}
where $\bm{\mathsf{\Psi}}^\dagger \bm{x}$ represents wavelet coefficients, $\| \cdot \|_{\ell_1}$ is the sum of absolute values, and $\|\cdot \|_{\ell_2}$ is the Euclidean norm.
\textit{Maximum a posteriori} (MAP) estimation is found by choosing the estimate of $\bm{x}$ that will maximize the posterior,
which is equivalent to minimizing the negative log posterior, i.e.
\begin{equation}  \label{eq:MAP-estimate}
\begin{aligned}
  {\rm arg}\max_{\bm{x}} p\left(\bm{x} | \bm{y} \right) 
    &=\\ {\rm arg}\min_{\bm{x}} &\left\{ \|\bm{y} - \mathsf{\bm{\Phi}} \bm{x}\|_{\ell_2}^2/2\sigma^2 + \gamma \| \bm{\mathsf{\Psi}}^\dagger \bm{x}\|_{\ell_1} \right\}.
\end{aligned}
\end{equation}
This minimization problem is known as regularized least squares, with the regularization term in the analysis setting being $\| \bm{\mathsf{\Psi}}^\dagger \bm{x}\|_{\ell_1}$. In many cases $\| \bm{\mathsf{\Psi}}^\dagger \bm{x}\|_{\ell_1}$ 
is chosen to penalize the number of parameters that determine $\bm{x}$ and reduce over fitting;
moreover, it can also be used to enforce other properties for $\bm{x}$ like smoothness. Furthermore, it is possible to add indicator functions as a prior that can restrict our solution to be real or positive valued, as is done in the constrained problem below. MAP estimation can be solved efficiently using the Forward Backward Splitting algorithm (\emph{e.g.} \cite{cai17b}).

An issue of using MAP estimation to perform sparse regularization is choosing a proper regularization parameter $\gamma$ (although there are ways to address this; \cite{per15}). The choice of $\gamma$, however, can be avoided after moving from the unconstrained problem in MAP estimation to the constrained problem
\begin{equation} \label{eq:model-con}
  {\rm arg}\min_{\bm x} \| \bm{\mathsf{\Psi}}^\dagger \bm{x}\|_{\ell_1} + \iota_{\mathcal{B}^\epsilon(\bm{y})}(\mathsf{\bm{\Phi}} \bm{x}) +\iota_{\mathbb{R}_+^N}(\bm{x})\, ,
\end{equation}
where $\iota$ is the indicator function that restricts $\mathsf{\bm{\Phi}} \bm{x}$ to the set \mbox{$\mathcal{B}^\epsilon(\bm{y}) = \{\bm{q} : \|\bm{y} - \bm{q} \|_{\ell_2} \leq \epsilon \}$},
 $\epsilon$ is the error tolerance, and $\iota_{\mathbb{R}_+^N}$ restricts the solution to be positive.

\subsection{ADMM}
\label{sec:admm}
In the constrained problem the $\ell_2$-norm is replaced with an indicator function that restricts the solution to the $\ell_2$-ball of radius $\epsilon$. This indicator function is non-differentiable and it is not possible to apply the Forward Backward method to obtain a solution.

Let $f(\mathsf{\bm{\Phi}} \bm{x}) = \iota_{\mathcal{B}^\epsilon(\bm{y})}(\mathsf{\bm{\Phi}} \bm{x}) $ and $g(\bm{x}) = \| \bm{\mathsf{\Psi}}^\dagger \bm{x}\|_{\ell_1} +\iota_{\mathbb{R}_+^N}(\bm{x})$. We can then consider the optimization problem
\begin{equation} \label{eqn:model-general-L}
  \min_{\bm{x} \in \mathbb{R}^N} f(\mathsf{\bm{\Phi}} \bm{x}) + g(\bm{x})\, .
\end{equation}
Problem \eqref{eqn:model-general-L} can be addressed by the alternating direction method of multipliers (ADMM) algorithm \cite{boy11,yan11,ono16}. Starting from an augmented Lagrangian, variables $\bm{x}$ and $\bm{v}$ are minimized alternatively while updating the dual variable 
$\bm{z}$ (using the dual accent method; \cite{boy11}) to ensure that the constraint $\bm{v} = \mathsf{\bm{\Phi}}\bm{x}$ is met in the final solution, i.e. 
\begin{align}
    \bm{x}^{(k)} & = \argmin_{\bm{x}\in \mathbb{R}^N} \left(g(\bm{x}) + \frac{1}{2\lambda}\|\mathsf{\bm{\Phi}}\bm{x} - (\bm{v}^{(k)} - \bm{z}^{(k)})\|^2_{\ell_2}\right) \label{alg:admm-x}\\
    \bm{v}^{(k+1)} & = \argmin_{\bm{v}\in \mathbb{R}^K} \left(f(\bm{v}) + \frac{1}{2\lambda}\|\bm{v} -(\mathsf{\bm{\Phi}}\bm{x}^{(k)} + \bm{z}^{(k)})\|^2_{\ell_2}\right) \label{alg:admm-v} \\
    {\bm{z}}^{(k+1)} & = \bm{z}^{(k)} +  (\mathsf{\bm{\Phi}}\bm{x}^{(k)} - \bm{v}^{(k+1)})\, .
\end{align}
The above iterations can be evaluated using a combination of proximal operators and the Dual Forward Backward algorithm (allowing the use of more than one wavelet basis). See \cite{pra19c} for a detailed description of the ADMM algorithm applied here.

\section{Reconstructions}
\label{sec:reconstructions}
In this section we demonstrate reconstruction of simulated SPIDER observations using the ADMM algorithm, where a solution is found from the constrained problem. We use the software package PURIFY\footnote{\url{https://github.com/astro-informatics/purify}} to perform interferometric image reconstruction, powered by the convex optimization package SOPT\footnote{\url{https://github.com/astro-informatics/sopt}}.

To generate the measurement operator used to simulate the observation we use the Kaiser-Bessel kernel with a support size $J = 8$ pixels to reduce aliasing error in the ground truth measurements. For reconstruction, we use a measurement operator with a kernel support size of $J = 4$ pixels. The number of pixels in $\bm{x}$ are determined by the ground truth image, $\bm{x}_{\rm Ground Truth} \in \mathbb{R}^N_+$. We do not include the decrease in sensitivity of the SPIDER instrument away from the center of the field, but this can be included in simulations if it is well characterized. To simulate the observation we follow \cite{pra18} and add i.i.d.\  Gaussian noise to the observational data. We define an input signal to noise ratio (ISNR) to determine the standard deviation of the Gaussian noise, where this standard deviation is defined as
\begin{equation}
  \sigma_i = \frac{\| \mathsf{\bm{\Phi}} \bm{x}_{\rm Ground Truth}\|_{\ell_2}}{\sqrt{M}} \times 10^{-\frac{\rm ISNR}{20}}\, .
\end{equation}
The Fourier sampling pattern of the observation (i.e. the $uv$-coverage) is determined by the design of the SPIDER instrument and the optical spectral coverage. By combining the entire spectra it is possible to increase the sampling coverage, as explained in Section \ref{sec:spider}. We use the configuration of Table \ref{tab:spider_specs} (shown in Figure \ref{fig:uvcoverage}).

\begin{figure}
    \center
    \includegraphics[width=0.45\textwidth]{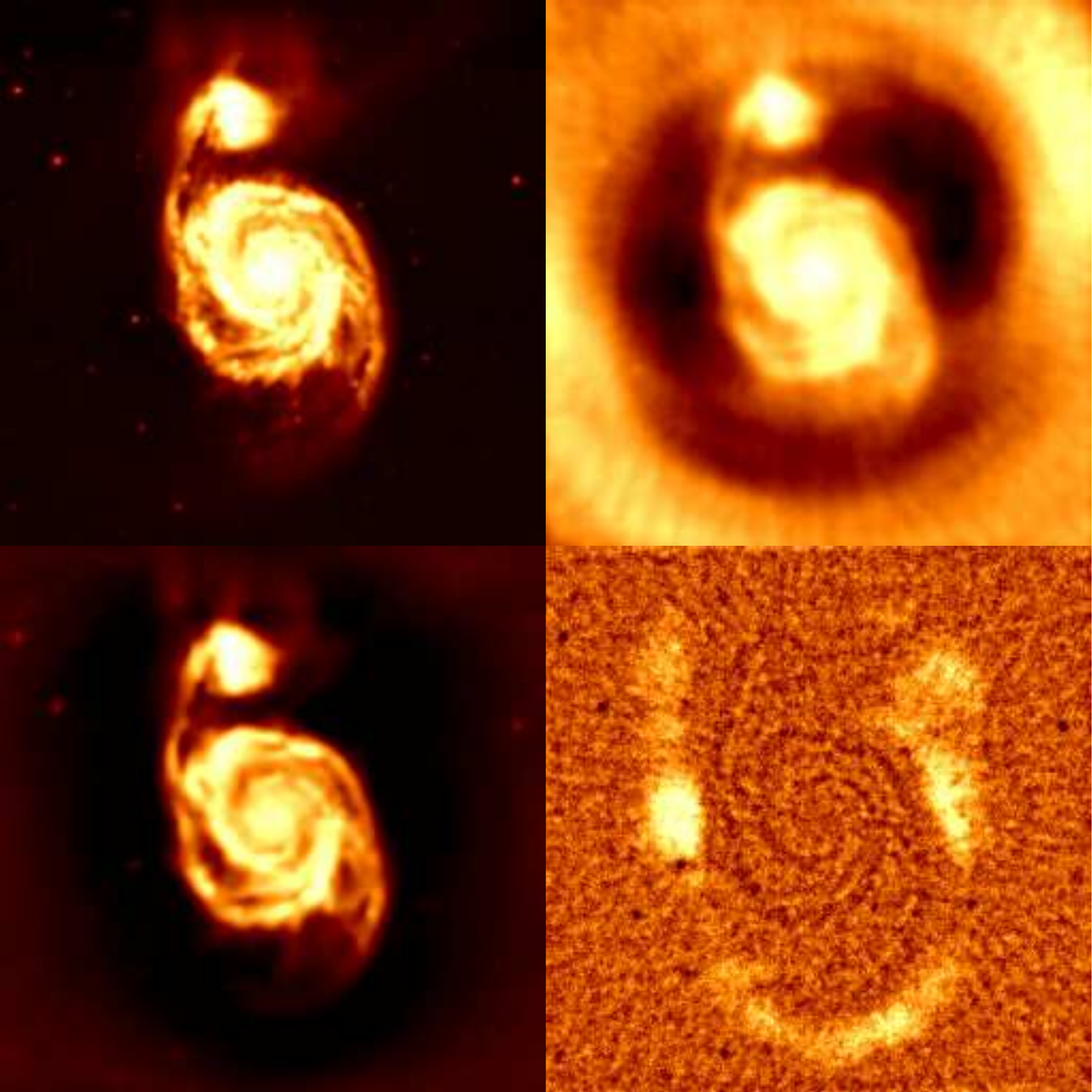}
    \caption{Simulation of observation and reconstruction of the spiral galaxy M51 using ADMM implemented with PURIFY, including the ground truth (top left), the observed image (top right), the PURIFY reconstruction (bottom left), and the residuals (bottom right). We used an ISNR of 30dB, a pixel size of 0.3$^{\prime\prime}$, and an image size of 256 by 256 pixels, with the sampling pattern for 10 spectral bins as shown in Figure \ref{fig:uvcoverage} resulting in 4440 measurements. The structure of the spiral arms and point sources are recovered well using PURIFY.}
    \label{fig:reconstructions}
\end{figure}

The results presented in Figure \ref{fig:reconstructions} show that an observation using the proposed SPIDER design can be effectively reconstructed using PURIFY. Reconstruction was performed using a Dirac basis and Daubechies wavelets 1 to 8. While we have used the design from \cite{ken13}, where the baselines lie on radial spokes, different baseline configurations may lead to higher quality reconstruction. Depending on the structures in the ground truth sky, different baseline configurations will be more effective at sampling the sky, leading to more effective reconstruction of objects and their details. It was recently shown that the theory of compressive sensing might lead to more efficient designs \cite{liu18}.

In summary, we adapt recent developments in radio interferometric imaging, leveraging sparsity and convex optimisation, and show that they are effective for imaging SPIDER observations.  Moreover, recent developments in efficient uncertainty quantification for radio interferometric imaging can also be adapted for use with SPIDER \cite{cai17b}. The computational performance of these algorithms can be further increased using GPU multi-threading and distribution across nodes of a computing cluster (as implemented in PURIFY already; \cite{pra19c}).

\vfill

\bibliographystyle{apalike}
{\small
\bibliography{example}}

\end{document}